\documentclass[final,1p,times]{elsarticle} 
\usepackage{graphicx} 
\usepackage{amssymb} 
\usepackage{amsthm} 
\usepackage{lineno} 

\journal{Nuclear Physics A} 
\begin{document} 

\begin{frontmatter} 


\title{Results from the commissioning of the ALICE Inner Tracking System with cosmics}

\author{Francesco Prino$^{a}$ for the ALICE collaboration}

\address[a]{INFN, 
Sezione di Torino,
Italy}

\begin{abstract} 
The Inner Tracking System (ITS) is the detector of the ALICE
central barrel located closest to the beam axis and it is
therefore a key detector for tracking and vertexing performance.
Here, the main results from the ITS  
commissioning with atmospheric muons in 2008 are presented, focusing
in particular on the detector operation and calibration and on
the methods developed for the alignment of the 
ITS detectors using reconstructed tracks.
\end{abstract} 

\end{frontmatter} 



\section{Introduction}

The Inner Tracking System (ITS) of the ALICE experiment~\cite{TPAP} consists
of six cylindrical layers of silicon detectors located at radial 
distances of $\approx$ 3.9, 7.6, 15, 24, 38 and 43 cm from the beam axis and
covering the pseudo-rapidity range $|\eta|<$0.9.
The two innermost layers are equipped with 240 pixel detectors (SPD), 
the two intermediate layers are made of 260 drift detectors (SDD), while
1698 strip detectors (SSD) are mounted on the two outermost layers.
The number, position and segmentation of the layers, as well as the detector 
technologies, have been designed according to the requirements of 
efficient track finding in the high multiplicity environment predicted 
for central Pb--Pb collisions at LHC,
high resolution on track impact parameter 
and minimization of material budget (multiple scattering).
Hence, the ITS allows to improve the momentum and angle resolution for 
tracks reconstructed in the TPC, to recover 
particles that are missed by the TPC (due to either dead regions or 
low-$p_{\rm t}$ cut-off), to reconstruct the interaction vertex 
with a resolution better
than 100 $\mu$m and to identify the secondary vertices from the decay of
hyperons and heavy flavoured hadrons~\cite{PPR2}.
By associating pairs of reconstructed points in the two SPD
layers, it is possible to build ``tracklets'' that are used 
to reconstruct the interaction vertex position (which provides
a starting point for the Kalman filter in the tracking phase), to
tag pileup events on the basis of multiple vertices
and to measure the charged particle multiplicity in the wider 
pseudorapidity range ($|\eta| < 2$) covered by the SPD layers.
The four layers equipped with SDD and SSD provide also particle 
identification capability via dE/dx measurement.
This feature allows to use the ITS also as a standalone 
spectrometer, able to track and identify particles down to momenta
below 200 MeV/c.
Furthermore, the SPD FastOR digital pulses provide a unique 
prompt trigger signal.

\section{ITS installation and commissioning}

The installation of the ITS detectors and the beam pipe in the ALICE cavern 
was completed in June 2007.
A first commissioning run was performed in December 2007 to test the 
acquisition and the calibration strategy on a fraction of modules for
which power supplies and cooling were available.
A larger fraction of modules (about 1/2 of the full detector) participated
in the February/March 2008 data taking.
The installation of services was completed in May 2008.
The SPD FastOR was integrated in the Central 
Trigger Processor and since May 2008 it was used to collect cosmic data 
in self-triggering mode as well as to provide the trigger to other 
detectors. 
A data sample of more than 100k SPD-triggered atmospheric muons for first 
track-based alignment of the ITS sensors and charge signal calibration 
in SDD and SSD was collected in summer 2008.
The FastOR trigger for cosmic rays was given by a 
coincidence between the top and bottom half-barrels of the outer SPD 
layer and provided a trigger rate of $\approx$ 0.18 Hz.
The ITS standalone tracker~\cite{PPR2} was adapted to reconstruct the 
atmospheric muons as two back-to-back tracks
starting from a fake vertex along the muon trajectory built from the 
points of the two SPD layers.
A large number of sub-detector (SPD, SDD and SSD) specific calibration runs 
have also been collected to monitor the stability of the 
detector performance during 3 months of continuous operation.

\subsection{Detector calibration and operation}

{\bf SPD.} The detector performance was optimized tuning several 8-bit DACs 
integrated in the front-end electronics and taking dedicated calibration 
runs to verify the pixel response. 
The calibration procedure is fully integrated in the ALICE calibration 
framework. 
It is based on detector algorithms which determine the proper values of the 
DACs by analyzing data from dedicated runs using either the on chip test-pulse 
or particles crossing the detector. 
In particular, for each of the 1200 front-end chips the pixel-matrix 
response and the minimum threshold have been optimized in order to maximize 
the efficiency and minimize the number of noisy pixels. 
A typical threshold value is about 2800 electrons. 
The remaining individual noisy pixels, corresponding to less than 
0.15\%, are masked and the information is stored in the Offline Conditions 
DataBase (OCDB) to be used in the offline reconstruction.
The high power dissipation (1.32kW) mainly generated by the front-end 
chips requires an efficient cooling system to maintain an operating 
temperature around 30$^\circ$C. The performance of the C$_4$F$_{10} $ based 
evaporative cooling has been studied in detail; the cause of few spots of lower 
efficiency, which prevented the full operation of some modules, has been found 
and corrective action has been taken. 
Before the winter shutdown, 212 modules out of 240 could be stably 
operated for data taking and trigger purpose.
The SPD was operational during the LHC beam injection tests since June 2008 
and provided relevant information to study the level of background induced 
in ALICE by the LHC monitoring equipments. 

{\bf SDD.} 246 out of 260 SDD modules were included 
in DAQ during summer 2008. The SDD calibration strategy is based on three 
types of standalone
runs collected periodically during the data taking and analyzed 
by dedicated quasi-online algorithms that store the obtained calibration
parameters in the OCDB.
The first type is the pedestal run which 
allows to measure for each of the 133k anodes the values of
baseline and noise as well as to tag the noisy channels ($\approx$ 0.5\%).
The baselines are then equalized to a common value in the front-end analog
memory buffers~\cite{SDDfee}.
In the pulser runs, a test pulse is sent to the pre-amplifiers to measure 
the gain and to tag the dead channels ($\approx$ 1\%).
Finally, injector runs provide a measurement of the drift speed 
in 33 positions along the anode coordinate for each SDD module
by exploiting the MOS charge injectors integrated on the 
detector surface~\cite{SDDinj}.
This is a crucial element for the detector calibration since 
the drift speed depends on temperature (as $T^{-2.4}$) and it is 
therefore sensitive to temperature gradients in the SDD volumes and
to temperature variations with time.
A correction for non-uniformity
of the drift field (due to non-linearities in the voltage
divider and for few modules also to significant inhomogeneities in dopant
concentration) is applied: it 
is extracted from measurements of the systematic deviations 
between charge injection position and reconstructed coordinates that was 
performed on all the 260 SDD modules with an infrared laser~\cite{SDDmaps}.

{\bf SSD.} 1477 out of 1698 SSD modules took data in summer 2008. 
The fraction of bad strips was $\approx$1.5\%.
Most of the modules not included in the data taking were drawing unexpectedly 
high bias current and were switched off as a precaution, pending further 
investigation, although their performance was good. 
The SSD gain calibration has two components: overall calibration of ADC 
values to energy loss and relative calibration of the P and N sides. 
This charge matching is a strong point of double sided silicon sensors and 
helps to remove fake clusters. 
Both calibrations relied on cosmics. 
The calibration constants were determined already in the laboratory, 
using cosmics on a spare ladder. 
This was refined during the cosmics runs with the SPD FastOR trigger. 
The resulting normalized difference in P- and N-charge has a FWHM of 11\%. 
The absolute calibration matches within 5\% with the standard values for 
energy loss from the literature.

\subsection{Alignment with atmospheric muons}

A good knowledge of the real detector geometry is a key point
to obtain the design tracking performance of the ITS.
The displacements and deformations of the ITS modules 
should be reconstructed from dedicated samples of tracks 
(from both cosmic ray and p--p events) with a precision sufficient to limit 
the worsening of resolution due to misalignment at no more than 20\% of the 
nominal value. 
This is a challenging task having in mind that 
the total number of degrees of freedom 
is greater than 13000 and that
for SPDs
the nominal resolution on $r\varphi$ coordinate is 12 $\mu$m.
The relative position of ITS and TPC has also to be extracted with 
track-based methods and it is monitored by an 
an optical alignment system with an accuracy better than 20 $\mu$m~\cite{AMS}.
The starting point for the ITS internal alignment procedure are the optical 
measurements (survey) performed 
during the construction phase.
Then, track-based methods are applied following a hierarchical sequence
which, for each sub-detector barrel (SPD, SDD and SSD), starts
from assemblies of sensitive elements mounted on common mechanical supports 
and moves at each step to smaller and smaller structures down to the level of 
single modules.
Two distinct methods based on the minimization of the residuals between
tracks and reconstructed points are used, namely the Millepede~\cite{millepede} 
and a module-by-module iterative approach.
The results obtained with the iterative method are of similar quality
as the Millepede ones. A comparison of the extracted alignment parameters 
with the two methods can be used to assess the systematics. 

\begin{figure}[b!]
\vskip -0.3 cm
\resizebox{0.33\textwidth}{!}
{\includegraphics*[]{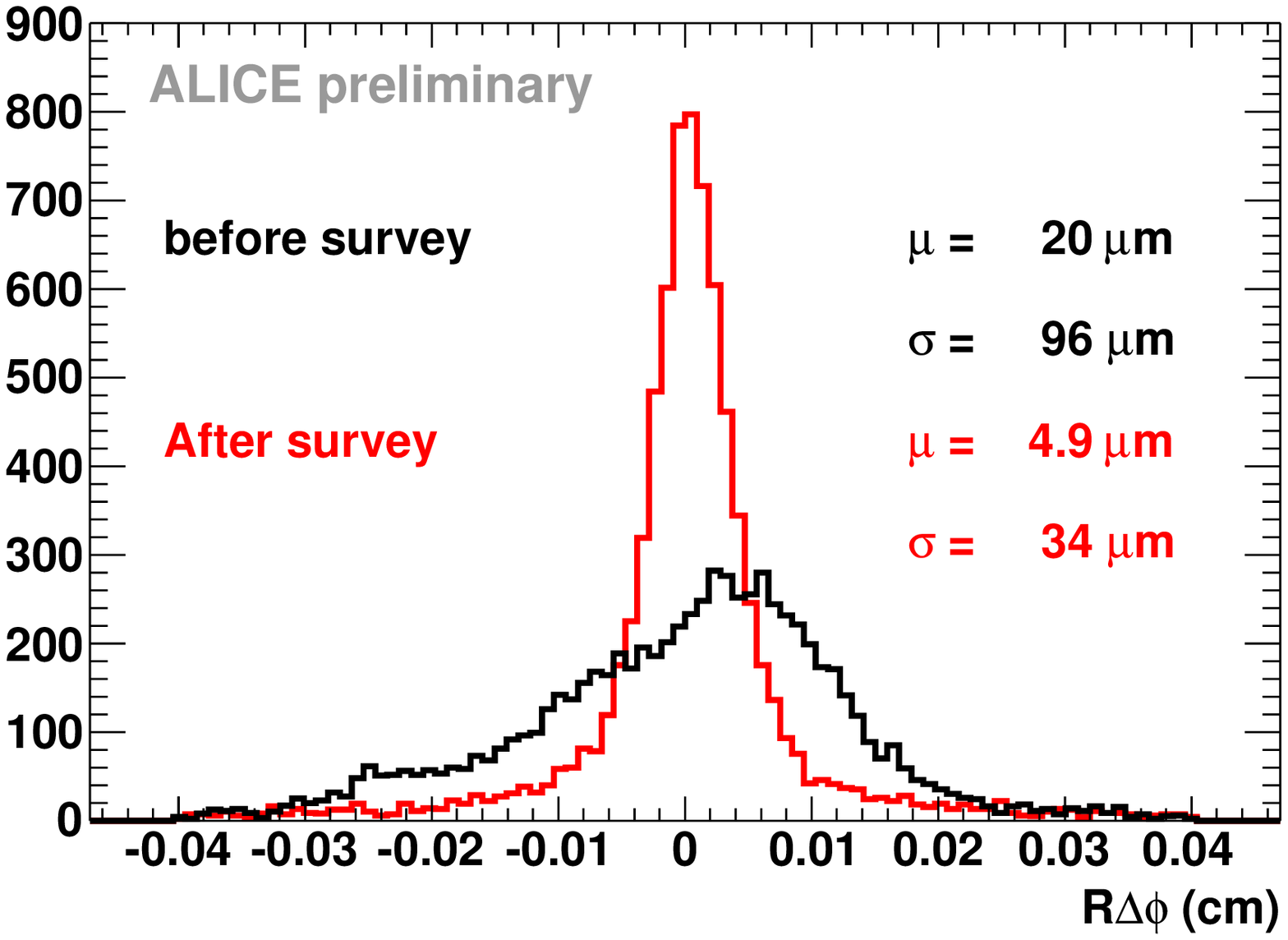}}
\resizebox{0.33\textwidth}{!}
{\includegraphics*[]{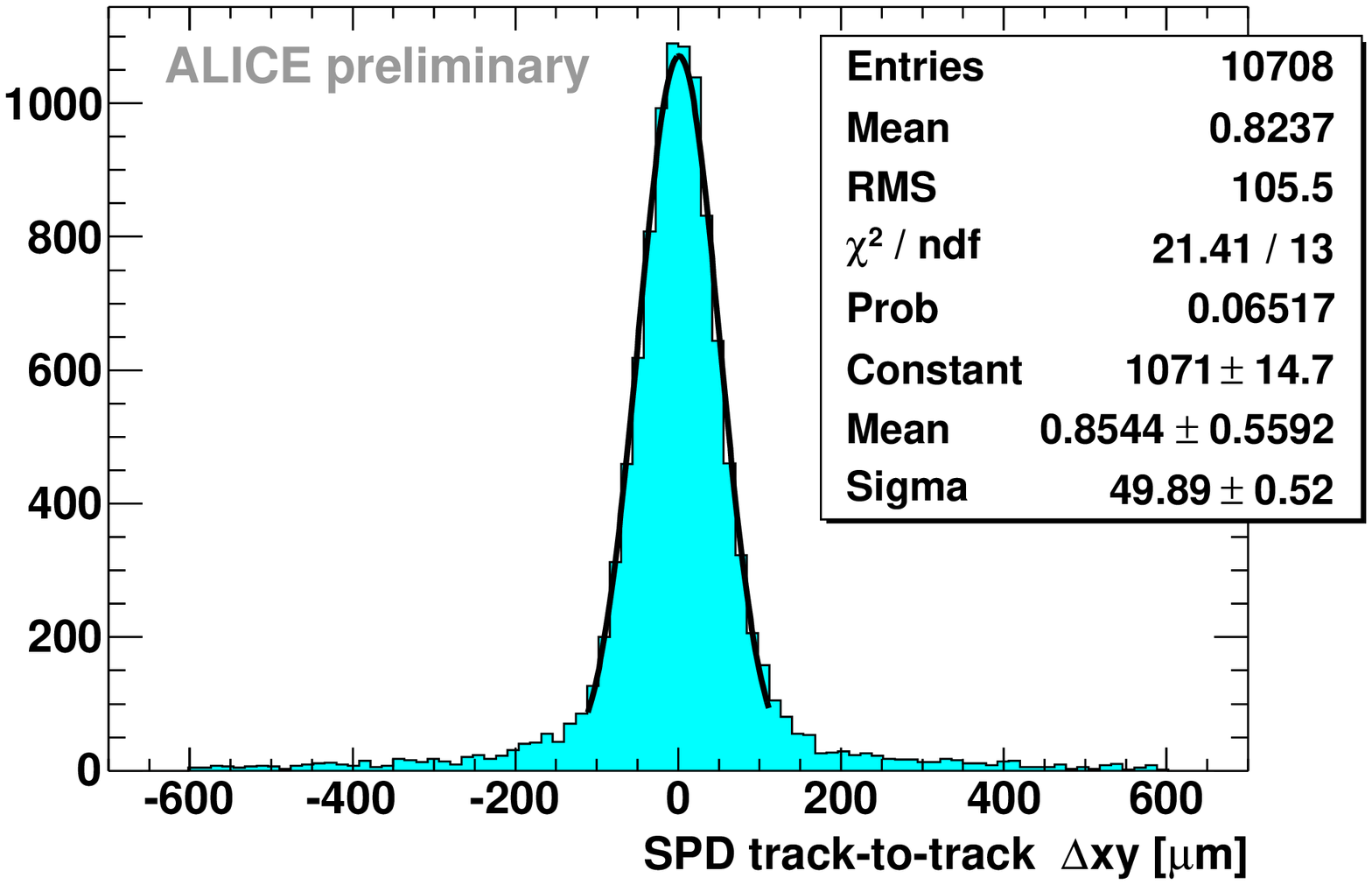}}
\resizebox{0.33\textwidth}{!}
{\includegraphics*[]{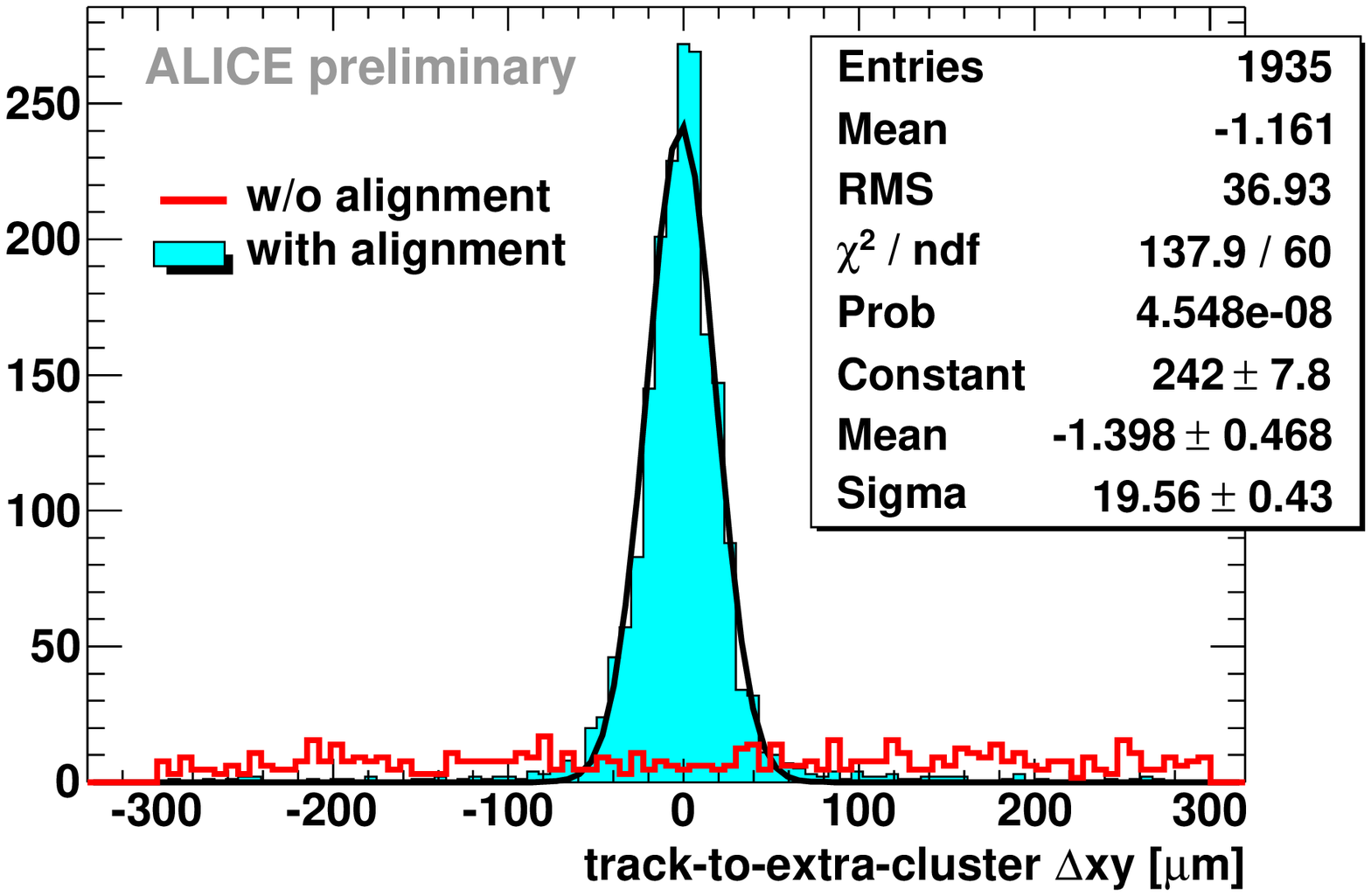}}
\vskip -0.3 cm
\caption{Left: track-to-point residuals in SSD with and without information from survey. Center: Track-to-track $\Delta$xy at y=0 after Millepede re-alignment.
Right: Track-to-extra-cluster $\Delta$xy before and after re-alignment.} 
\label{fig:ITSalign}
\end{figure}

After running the alignment tools, three main observables are used to check 
the quality of the obtained results.
First, track-to-point residuals are checked.
An example is shown in fig.~\ref{fig:ITSalign}-left where the distances 
in r$\varphi$ between the track fitted on the outer SSD layer and the 
points measured in the inner SSD layer are used to test the effect of the 
application of the SSD survey.
The second tool sensitive to alignment quality is the comparison 
of track segment parameters (inclination and position) after splitting
a muon track into two track segments crossing the top and bottom 
half-barrels respectively.
As an example, in the middle panel of fig.~\ref{fig:ITSalign} 
the distribution of the distance in the transverse plane ($\Delta$xy) 
for track segments reconstructed in the 
top and bottom part of SPD layers and propagated to y=0 is reported. 
It can be seen that after Millepede re-alignment a Gaussian 
r.m.s. of residual distributions of 50$\mu$m is obtained, which should be 
compared with the 40$\mu$m that are obtained from detailed Monte Carlo 
simulations of the ALICE apparatus with ideal geometry.
Finally, the third used observable is the distance between 
clusters in the region where there is an acceptance overlap between two
adjacent modules (extra-clusters). 
A particle passing through these overlapping regions produces
two points very close in space whose reconstructed distance is sensitive 
to the relative misalignment of the two modules. 
In fig.~\ref{fig:ITSalign}-right the track-to-point distance for the 
SPD extra-clusters in the $r\varphi$ plane before and after Millepede 
re-alignment is displayed.
From the width of the distribution, one can estimate the spatial 
resolution on the single SPD point as 
$\sigma_{spatial}\approx \sigma_{\Delta\rm{xy}}/\sqrt{2}\approx 14~\mu$m, which has to 
be compared with the 11 $\mu$m that are obtained from simulations with 
ideal geometry. 
The alignment of SDD detectors
for the r$\varphi$ coordinate (reconstructed from the drift time)
is complicated by the interplay between the geometrical misalignment 
and the calibration of drift speed and $t_0$ 
(i.e. the measured drift time for particles with zero drift 
distance).
The $t_0$ can be extracted either from the minimum measured drift 
time or from the track-to-point residual distributions along the drift 
direction.
These distributions, in case of mis-calibrated $t_0$, show
two opposite-signed peaks due to the presence, in each module, of two 
separated drift regions where electrons move in opposite directions.
After a first calibration with these methods, a refinement is obtained 
by adding in the Millepede the $t_0$ as a free global 
parameter for each of the 260 SDD modules.
Similarly, the drift speed has been added as a free parameter 
for those SDD modules with mal-functioning injectors.

\subsection{Charge calibration with atmospheric muons}

The atmospheric muons provide also a sample of 
ionizing particles for absolute calibration of the dE/dx 
(from ADC units to keV) for SDD and SSD.
In case of SDD detectors, the collected charge actually depends on
drift distance due to the applied zero-suppression:
the larger the drift distance, the larger the charge diffusion and
consequently the charge cluster develops wider tails which get more
easily cut by the zero-suppression algorithm.
The data sample of atmospheric muons allowed to validate 
the correction for this effect extracted from detailed 
simulations of the detector response.
The correction for track inclination has also been tested 
and, for a small sample of tracks collected with magnetic field, a
comparison of the SSD signals with the Bethe-Bloch expectation 
values has been done.
All these checks are also important to validate the detector
responses implemented in the simulations and used in the particle 
identification algorithms.

\section{Conclusions}

A successful data taking for commissioning the ALICE ITS has been
carried out in summer 2008 with SPD-triggered atmospheric muons.
It allowed to obtain a good knowledge of the calibration parameters for 
all the detectors 
and a first alignment for 85\% of SPD and 50\% of SSD modules.
The ITS results to be well performing and ready for 
p--p and Pb--Pb collisions.

\end{document}